\documentclass[11pt]{article}

\usepackage{amsmath}
\usepackage{amssymb}
\usepackage{lineno}

\usepackage{graphicx}
\usepackage{epstopdf}
\usepackage[round]{natbib}
\usepackage{color, cancel}

\usepackage{setspace} 
\usepackage[labelfont=bf,labelsep=period,justification=raggedright]{caption}

\makeatletter
\renewcommand{\@biblabel}[1]{\quad#1.}
\makeatother

\date{}

\makeatletter
\def\tagform@#1{\maketag@@@{\ignorespaces#1\unskip\@@italiccorr}}
\makeatother

\begin{document}
\bibliographystyle{plainnat}

\begin{flushleft}
{\Large
\textbf{A generalized residual technique for analyzing complex movement models using earth mover's distance}
}
\\
{\footnotesize
Jonathan R. Potts$^{1,a}$, 
Marie Auger-M\'eth\'e$^{1,2,b}$
Karl Mokross$^{3,4,c}$
Mark A. Lewis$^{1,2,d}$
}
\end{flushleft}
\vspace{5mm}\noindent\textbf{Short title:} Generalized residual technique


\vspace{10mm}

\begin{flushleft}
{\footnotesize
\bf{1} Centre for Mathematical Biology, Department of Mathematical and Statistical Sciences, University of Alberta, Canada
\\
\bf{2} Department of Biological Sciences, University of Alberta, Edmonton, Canada
\\
\bf{3} {School of Renewable Natural Resources, Louisiana State University Agricultural Center, Baton Rouge, Louisiana, 70803}
\\
\bf{4} {Projeto Din\^amica Biol\'ogica de Fragmentos Florestais. INPA. Av. Andr\'e Ara\'ujo 2936. Petr\'opolis. Manaus. Brazil. 69083-000}
\\
\bf{a} E-mail: jrpotts@ualberta.ca.  Centre for Mathematical Biology, Department of Mathematical and Statistical Sciences, 632 CAB, University of Alberta, Canada, T6G 2G1.  Tel: +1-780-492-1636.

\bf{b} E-mail: marie.auger-methe@ualberta.ca

\bf{c} E-mail: kmokro1@tigers.lsu.edu.  

\bf{d} E-mail: mark.lewis@ualberta.ca.  
}
\end{flushleft}

\newpage
\section*{Summary}

1.  Complex systems of moving and interacting objects are ubiquitous in the natural and social sciences.  Predicting their behavior often requires models that mimic these systems with sufficient accuracy, while accounting for their inherent stochasticity.  Though tools exist to determine which of a set of candidate models is best relative to the others, there is currently no generic goodness-of-fit framework for testing how close the best model is to the real complex stochastic system.  

{\noindent}2. We propose such a framework, using a novel application of the {\it Earth mover's distance}, also known as the {\it Wasserstein metric}.  It is applicable to any stochastic process where the probability of the model's state at time $t$ is a function of the state at previous times.  It generalizes the concept of a residual, often used to analyze 1D summary statistics, to situations where the complexity of the underlying model's probability distribution makes standard residual analysis too imprecise for practical use.

{\noindent}3. We give a scheme for testing the hypothesis that a model is an accurate description of a data set.  We demonstrate the tractability and usefulness of our approach by application to animal movement models in complex, heterogeneous environments.  We detail methods for visualizing results and extracting a variety of information on a given model's quality, such as whether there is any inherent bias in the model, or in which situations it is most accurate.  We demonstrate our techniques by application to data on multi-species flocks of insectivore birds in the Amazon rainforest.

{\noindent}4. This work provides a usable toolkit to assess the quality of generic movement models of complex systems, in an absolute rather than a relative sense.

\newpage
\section*{Introduction}

How good is a model at describing reality?  This fundamental question, ubiquitous across the quantitative sciences, has troubled and intrigued scientists for over 200 years \citep{legendre1805, gauss1809}.  A variety of techniques have been discovered to address the problem in certain situations.  Residual analysis is one example that has a long history of useful application in various areas \citep{zuuretal2009, gordon2010}.  However, it is only usable when the underlying model, or a summary statistic arising from the model, can be framed as a simple deterministic function.

Despite this, our world is infused with complex, multi-dimensional, stochastic systems.  These range from biological systems, such as ant colonies, bird flocks and slime mold aggregation \citep{camazineetal2003}, to crowd movement psychology in social sciences \citep{helbingetal2007}, to protein dynamics \citep{berendsenhayward2000}.  Such systems are typically high-dimensional and can rarely be described in an accurate way without taking into account underlying randomness in movements of constituent objects.  The aim of this paper is to generalize the technique of residual analysis so that it can be used for generic stochastic systems of moving and interacting objects.

The type of models that are analyzable by residual analysis can be characterized as {\it deterministic} models.  These are models of the form ${\bf a}=f({\bf b})$, where ${\bf a}$ is the prediction, ${\bf b}$ is a vector of independent input variables and $f$ is a deterministic function.  The {\it residual}, ${\bf a}_{\rm obs}-f({\bf b})$, where ${\bf a}_{\rm obs}$ is an observation, measures the closeness of the model to the data.  Residual analysis is well-developed and often used for assessing the quality of models arising from techniques such as regression \citep{zuuretal2009,gordon2010}.  However, when the function $f$ is replaced by a probability distribution, $P({\bf b})$, residuals are no longer well-defined.  If the distribution is sufficiently close to a Gaussian, such as if $P({\bf b})=f({\bf b})+\xi$ where $\xi$ is a zero-mean noise term and $f$ is deterministic, one can simply define the residual to be the distance between the data point ${\bf a}_{\rm obs}$ and the mean of $P({\bf b})$.  However, this fails to be reasonable if the distribution is more complex, for example multimodal or long tailed.


Typical stochastic movement-and-interaction models often depend on heterogeneous properties of either the environment \citep{foresteretal2009, vanmoorteretal2009, PBMSL} or surrounding agents \citep{camazineetal2003}, frequently making the probability distribution of state transitions complex and multi-peaked. 
While methods exist for selecting the {\it relative} quality between competing models of these complex systems, such as Likelihood Ratio \citep{PBMSL}, Akaike Information Criteria (AIC), Deviance Information Criteria \citep{moralesetal2004} or Bayesian methods \citep{jonsenetal2005}, the current suite of goodness-of-fit tests fail to provide sufficient techniques for assessing the {\it absolute} quality of such a model: that is, its closeness to the data.  This has led to researchers either ignoring the question and solely performing model selection \citep{moorcroftetal2006}, or performing {\it ad hoc} tests on 1D summary statistics \citep{grimmetal2005, gautestadetal2013}.  For example, a search for the 20 highest cited papers that fit animal movement models to data reveals that {\it none} test the absolute fit of the best model to the data (methods in SI Appendix C).

In the animal movement literature in particular, this tendency to ignore the absolute quality of a model has been partially responsible for various controversies regarding the detection of underlying movement processes \citep{augermetheetal2011}.  This has led to criticism of many papers for appearing to draw strong conclusions about animal behavior by selecting the best of a small number of simple models, all of which may be very poor at reflecting data.  For example, the results of \citet{viswanathanetal1996} were later overturned by \citet{edwardsetal2007}, and \citet{dejageretal2011} was criticized by \citet{jansenetal2012} for drawing possibly incorrect conclusions by only examining very simplistic models.  

Recent work \citep{augermetheetal2011, augermetheetal2014} demonstrates that these issues may sometimes be resolved by examining the residuals of the respective models' step length and turning angle distributions.  However, this technique is only applicable to a specific set of models, which have relatively simple distributions, and cannot easily incorporate the effects of heterogeneous surroundings on movements.  Increasingly, it is proving necessary to factor such effects into movement models.  Recent developments in both the step selection literature \citep{foresteretal2009, fortinetal2005, rhodesetal2005, latombeetal2013, vanakaetal2013, PBMSL} and collective behavioral studies \citep{deneubourgetal1989, couzinetal2002, hoareetal2004, guttalcouzin2010} amply demonstrate the importance of incorporating often heterogeneous surroundings into the understanding and modeling of animal movement.  It is therefore necessary to construct tools similar to residual analysis, yet applicable to these more complex models, to avoid repeating the sort of problems that have already plagued the field of movement ecology regarding simpler models, often caused by choosing between a limited set of possibly poor models \citep{plankcodling2009, dejageretal2011, gautestadetal2013}.

With these issues in mind, we construct a generalized residual method that can be applied to any stochastic system of moving and interacting objects.  The particular types of models that we are concerned with are one-step Markovian, describing the probability $P_\tau({\bf X}_{t+\tau}|{\bf X}_t)$ of a system being in state ${\bf X}_{t+\tau}$ at time $t+\tau$ having been in state ${\bf X}_{t}$ at time $t$, as this eases notation and explanation.  However, generalizing to non-Markovian situations merely involves re-writing the probability function so that it is dependent on several previous steps rather than just one.  The states ${\bf X}_{t+\tau}$ and ${\bf X}_{t}$ could consist of a variety of information about the system, for example the positions of the agents, their directions, environmental information perceived by the agents and so forth; whatever is appropriate for the scientific questions being addressed.

Large classes of complex systems models in the movement literature can be described in this way, as so-called {\it coupled step selection functions} \citep{PML}.  These include models of collective behavior, which are often applied to both human systems and inter- and intra-cellular systems \citep{berendsenhayward2000, camazineetal2003, helbingetal2005}.  Therefore our technique fills a gap in the increasingly important field of complex systems science, important for ecological applications and beyond.

\subsection*{Earth mover's distance: a measure of absolute fit}

Suppose that a complex system is described in `reality' by a function $P^R({\bf X}_{t+\tau}|{\bf X}_{t})$ but that the best model of the system is given by $P^M({\bf X}_{t+\tau}|{\bf X}_{t})$.  In other words, if the system is currently in state ${\bf X}_{t}$ then the probability distribution function of it being at state ${\bf X}_{t+\tau}$ after a time of $\tau$ has elapsed is $P^R({\bf X}_{t+\tau}|{\bf X}_{t})$.  However, the best model constructed so far predicts that the system will have probability distribution function $P^M({\bf X}_{t+\tau}|{\bf X}_{t})$.  To assess how well this model reflects reality requires a measure of the distance between the two probability functions $P^R$ and $P^M$.  (Note that $P^R$ and $P^M$ depend upon the time interval $\tau$ between successive states of the system, but $\tau$ remains fixed throughout the paper so we do not include it in the notation.)

\begin{figure}[ht!]
\includegraphics[width=150mm]{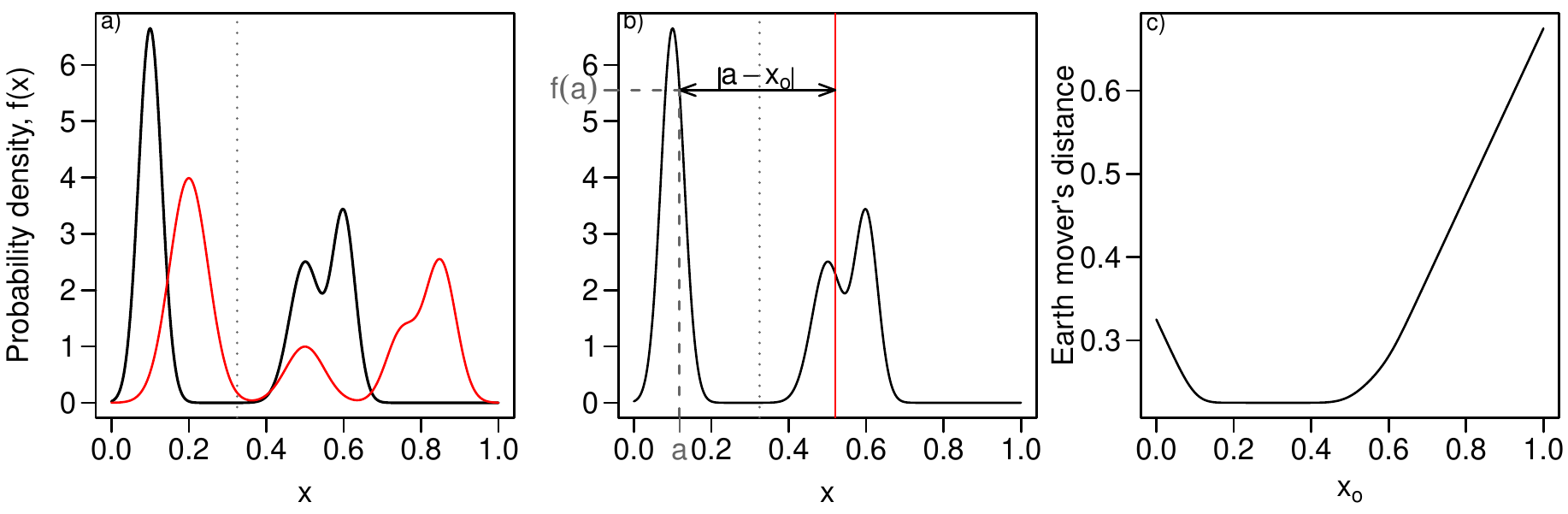}
\caption{\small {\bf The Earth Mover's Distance (EMD).} Panel (a) shows two probability distributions.  Imagine that the black one is a pile of earth and we want to construct from this the red distribution in the most efficient way possible.  The EMD gives the average distance that we would need to move each particle of earth when performing this transformation.  Panel (b) shows the situation relevant to the present study, where we have a model probability distribution of possible states that a complex system might be in at time $\tau$ in the future (black curve), together with an observation of the state at which the system actually ended up in, denoted by point $x=x_o$.  This observation translates to a Dirac delta probability density function $\delta(x-x_o)$, assuing that the observation has negligible error.  The EMD in this instance is the average distance each part of the probability density function has to move to end up at $x_o$.  For example, at point $x=a$, we need to move $f(a)=5.5$ amount of probability distribution a distance of $|a-x_o|$.  By integrating the product $|a-x_o|f(a)$ over all such $a$, we obtain the EMD between the model and data, for a single data point (\ref{EMD_single}).  The dotted line denotes the mean of the black distribution. Though we illustrate this in 1 dimension for ease of explanation, typically complex movement models may have states in much higher dimensions.  Panel (c) shows the EMD as a function of the observation $x_o$.}
\label{emm}
\end{figure}

Mathematicians have developed such a distance function, called the {\it Wasserstein metric} \citep{wasserstein1969}, a special case of which has recently re-emerged in the visual biometrics literature as the {\it earth mover's distance} \citep{rubneretal2000}.  Though the general measure-theoretic definition is rather formal and technical (SI Appendix B), the distance has an intuitive explanation.  Imagine that one of the probability distributions describes a pile of earth (e.g. sand, soil, etc.) that you have in front of you, and the other the describes the shape of a pile of earth that you want to construct.  Intuitively, the earth mover's distance is the minimum average distance that each particle of earth has to move in order to change the pile from what you have to what you want (Fig.~\ref{emm}a).

Though simple to state, this distance can be computationally complex due to an inherent minimization procedure \citep{pelewerman2009}.  However, in practice we are often interested in how close a movement model is to a {\it data set}, rather than a probability distribution that reflects {\it reality}.  
It turns out that the earth mover's distance between a model and a data set is considerably easier to compute than that between two probability distributions, as it obviates the need for minimization (see SI Appendix B).  

Suppose we have data on a complex system saying that it is in states ${\bf S}_0,{\bf S}_1,\dots,{\bf S}_N$ at times $0,\tau,2\tau,\dots,N\tau$ respectively.  Then the probability density function describing the transition between data point $n-1$ and $n$ is just a Dirac delta function $P^R({\bf X}|{\bf S}_{n-1}) = \delta({\bf X}-{\bf S}_{n})$.  In other words the probability of the system transitioning to any state other than ${\bf S}_{n}$ is zero, and the integral of the probability density function is equal to 1.  

Suppose also that the best model we have so-far constructed for these data is $P^M({\bf X}_{t+\tau}|{\bf X}_{t})$.  Then the {\it earth mover's distance} (EMD) between this model and a data point ${\bf S}_n$, given a previous data point ${\bf S}_{n-1}$, is
\begin{linenomath*}
\begin{align}
\mbox{EMD}(P^M;{\bf S}_n) &= \int_\Omega d({\bf X},{\bf S}_{n}) P^M({\bf X}|{\bf S}_{n-1}) {\rm d}{\bf X},
\label{EMD_single}
\end{align}
\end{linenomath*}
where $d$ is a distance metric between the states of the system and $\Omega$ is the space of all system states.  For example, $d$ could be the Euclidean distance $D_E$ between two points in space and $\Omega$ could be a subset of a 2-dimensional plane, if modeling a single terrestrial animal's movement.  As another example, for a collective system with $K$ animals, $d$ could represent the mean Euclidean distance between pairs of points for each animal, $d({\bf x}^1,\dots,{\bf x}^K|{\bf y}^1,\dots,{\bf y}^K)=\frac{1}{K}\sum_{k=1}^K D_E({\bf x}^k,{\bf y}^k)$, where ${\bf x}^k,{\bf y}^k$ are points in $\Omega$.   If the model is in discrete space, so that $\Omega$ is a finite set of points, one simply replaces the integral in \ref{EMD_single} with a sum and divide by the number of points in $\Omega$.  
An illustration of~\ref{EMD_single}, in the simplest case of one agent moving in one dimension, is given in Fig.~\ref{emm}b.  Notice that the Kullback-Liebler distance [see e.g. \citet{burnhamanderson}] only gives information based on the value of $f(x_o)$, whereas EMD takes into account the shape of the entire distribution $f(x)$.   

The definition in \ref{EMD_single} implicitly assumes that the noise in the data is negligible.  This is often reasonable for movement models constructed from GPS data of animals, since the error in GPS trackers is very highly correlated \citep{severnsbreed2014}.  However, if it is necessary to take into account of noise, this can be done by replacing \ref{EMD_single} with the general definition of EMD, given in SI Appendix B equation (1) with $p=1$.  Though analytically simple, the EMD in this case is much more intensive to compute than \ref{EMD_single}.  See SI Appendix B for more details.

Notice that if the model were deterministic then the state ${\bf X}$ of the model at time $t+\tau$ given that it was at state ${\bf X}_{t}$ at time $t$ is ${\bf X}_{t+\tau}=f({\bf X}_{t})$ for some function $f$.  Writing this in the notation of probability distributions, we have $P^M({\bf X}_{t+\tau}|{\bf X}_{t}) = \delta[f({\bf X}_{t})]$ so that the earth mover's distance for each data point ${\bf S}_n$ is precisely the absolute residual $|{\bf S}_n-f({\bf S}_{n-1})|$ of the model ${\bf X}_{t+\tau}=f({\bf X}_{t})$.  In other words, \ref{EMD_single} generalizes the concept of a residual, rationalizing the choice of this particular metric over the others available \citep{gibbssu2002}.

The EMD between a model and the whole data set ${\bf S}_0,{\bf S}_1,\dots,{\bf S}_N$ is the mean of \ref{EMD_single} over all the data points
\begin{linenomath*}
\begin{align}
\mbox{EMD}(P^M; {\bf S}_0,\dots,{\bf S}_N) &= \frac{1}{N}\sum_{n=1}^{N}\mbox{EMD}(P^M;{\bf S}_n).
\label{EMD}
\end{align}  
\end{linenomath*}
One drawback of the EMD is that it gives more weight to distributions with higher variance.  Also, it is a dimensional quantity, with units of space.  To mitigate against these issues, we use the dimensionless {\it standardized EMD} (SEMD), $E_{\rm S}(P^M; {\bf S}_0,\dots,{\bf S}_N)$, which is defined by dividing the EMD by the standard deviation $s_n$ of the model.  For a single data point, this is
\begin{linenomath*}
\begin{align}
\mbox{EMD}_{\rm S}(P^M;{\bf S}_n) &= \frac{\mbox{EMD}(P^M;{\bf S}_n)}{s_n},
\label{SEMD_single}
\end{align}
\end{linenomath*}
and $s_n$ is the standard deviation of the model for moving from position ${\bf S}_{n-1}$.  In other words,
\begin{linenomath*}
\begin{align}
s_n^2=\left[\int_\Omega {\bf X}^2 P^M({\bf X}|{\bf S}_{n-1}) {\rm d}{\bf X}-\left(\int_\Omega {\bf X} P^M({\bf X}|{\bf S}_{n-1}) {\rm d}{\bf X}\right)^2\right].
\label{SEMD_var}
\end{align}
\end{linenomath*}
The SEMD between a model and a sequence of data points ${\bf S}_0,{\bf S}_1,\dots,{\bf S}_N$ is 
\begin{linenomath*}
\begin{align}
\mbox{EMD}_{\rm S}(P^M; {\bf S}_0,\dots,{\bf S}_N) &= \frac{1}{N}\sum_{n=1}^{N}\mbox{EMD}_S(P^M;{\bf S}_n),
\label{SEMD}
\end{align}
\end{linenomath*}
Any method described in this paper using EMD can equally be performed using SEMD, so we explain everything just using EMD, for simplicity.  However, as the Results show, either SEMD or EMD may be preferable depending on the situation.

Further information about the model can be gained by looking at the mean directions from the model to the data, given by $\hat{\bf v}_n={\bf v}_n/|{\bf v}_n|$ where 
\begin{linenomath*}
\begin{align}
{\bf v}_n = \int_\Omega ({\bf S}_{n}-{\bf X}) P^M({\bf X}|{\bf S}_{n-1}) {\rm d}{\bf X},
\label{direction_bias}
\end{align}
\end{linenomath*}
so that $\hat{\bf v}_n$ is a unit vector in the direction from the data point ${\bf S}_{n}$ to the mean of the distribution $P^M({\bf X}|{\bf S}_{n-1})$ that predicts where ${\bf S}_{n}$ is likely to be.  When the system state is given by positions on a 2D plane, this information can be visualized by plotting each line from the origin to the position given by $\hat{\bf v}_n\mbox{EMD}(P^M;{\bf S}_n)$, giving a {\it wagon wheel} of directional EMDs (Fig.~\ref{wheels}a,b).  However, for a large data set, this can be somewhat messy.  Instead, we bin the directions into eight equal sections, constructing what we call a {\it dharma wheel} (Fig.~\ref{wheels}c-f), for its resemblance to the Buddhist symbol for the noble eightfold path \citep{beer2005}.  The smaller the dharma wheel, the more accurate the model.  

Dharma wheels are examples of the classical concept of a polar area diagram \citep{friendly2008}.  The choice of eight sections is quite arbitrary and, depending on the situation, it may be valuable to use a different number.  Alternatively, one could obtain a smoother wheel by fitting the wagon wheel to a mixture of wrapped normal distributions.  However, for simplicity of explanation, we use eight sections throughout this paper.

Dharma wheels also detect bias in data (Fig.~\ref{wheels}e,f), in analogy with residual analysis for linear models \citep{zuuretal2009}.  In addition to binning by direction, insight can be gained by constructing histograms of EMD against specific properties of the system (see Results).

An important use of EMD is to investigate goodness of fit statistically, by testing the null hypothesis `the model could have given rise to the data' against the alternative that it fails in this regard.  We assume the modeler has already used some form of selection technique (e.g. AIC, BIC) to find and parametrize the best of the models so far considered.  Then the following sequence of steps enables the modeler to find out whether this best model reflects the data well:
\begin{enumerate}
  \item Suppose there are $N$ data points (henceforth {\it the data}) and a best candidate model (henceforth, {\it the model})
  \item Simulate the model for $N$ steps and repeat $M$ times, where $M$ is as big as is computationally feasible
  \item For each simulation, generate the EMDs between each of the simulated paths and the model used to simulate them, to give $M$ distances $\mbox{EMD}_1,\dots,\mbox{EMD}_M$
  \item Find the EMD between the data and the model,  $\mbox{EMD}_{\rm data}$
  \item We can then test whether $\mbox{EMD}_{\rm data}$ is likely to be a sample from the distribution given by $\mbox{EMD}_1,\dots,\mbox{EMD}_M$.  We use a 5\% significance level, so that if $\mbox{EMD}_{\rm data}$ is greater than the 97.5 percentile of $\mbox{EMD}_1,\dots,\mbox{EMD}_M$, or less than the 2.5 percentile then we reject the null hypothesis.
\end{enumerate}
Notice that Step 5 is precisely equivalent to testing whether the Bayesian $p$-value $\mbox{Prob}(\mbox{EMD}_i \neq \mbox{EMD}_{\rm data}|\mbox{the model})$ is less than 5\% \citep{agrestihitchcock2005}.

\section*{Methods}

To show the practicality of our approach, we demonstrate how to use the earth mover's distance (\ref{EMD}) for models of animal movement in heterogeneous environments.   We use a simulated data set to test the efficacy of our model, based on an animal moving in an environment with two resource layers in a 1000 by 1000 square lattice (Fig.~\ref{ssf}).  These can be thought of, for example, as Geographic Information System (GIS) layers or resource distributions \citep{bolstad2005}.  The layers are Gaussian random fields, generated by the R function \texttt{GaussRF()} from the \texttt{RandomFields} package \citep{randomfieldsref}, using the \texttt{exponential} model.  Both layers have mean $=0$, variance $=1$, and nugget $=0$.  Layer 1 has scale $=10$, so varies rapidly through space.  For the sake of intuition, this might be thought of as denoting the amount of food available throughout the terrain.  For Layer 2, scale $=1000$, thus varies much more slowly than layer 1.  This layer could represent the topography or another large geographical constraint to movement, for example.
Disregarding the effect of the layers, animals move as random walkers with exponentially distributed step lengths which have a mean length of $5$ lattice points.  Then the effect of the layers on the animal's movement follows the concept of a step selection function, so that the probability $f({\bf x}|{\bf y})$ of moving to position ${\bf x}$ from position ${\bf y}$ in a time interval $\tau$ is given by 
\renewcommand{\arraystretch}{0.4}
\begin{linenomath*}
\begin{align}
f({\bf x}|{\bf y}) = K\exp[\underbrace{-\lambda|{\bf x}-{\bf y}|}_{\text{\begin{tabular}{c}step length \\ distribution\end{tabular}}}\underbrace{+\alpha w_1({\bf x})+\beta w_2({\bf x})}_{\text{\begin{tabular}{c}effect of \\ resource layers\end{tabular}}}],
\label{gen_model}
\end{align}
\end{linenomath*}
where $\lambda = 1/5$, $w_i({\bf x})$ is a function taking the value of Layer $i$ at position ${\bf x}$, $\alpha$ and $\beta$ are the model parameters, and $K$ is a constant that ensures that the integral of $f$ with respect to ${\bf x}$ is 1, so that $f$ is a probability distribution.  This is a simple version of a step selection function, or movement kernel, often used for modelling animal movement [e.g. \citet{fortinetal2005, foresteretal2009, PML}].

\renewcommand{\arraystretch}{1}

We generate two different simulated data sets.  One is of 100 different animals, starting at random locations in the grid, for 1000 `steps' (by which we mean `movements between successive location fixes') each.  The other is of 10 animals, again with random starting points, simulated for 500 steps each.  This enables us to demonstrate the relative effectiveness of our methods as applied to different sizes of data set.  The simulated data has $\alpha=1.5$ and $\beta=10$ (\ref{gen_model}).
We analyzed the simulated data $\{{\bf x}_0,\dots,{\bf x}_N\}$ (which can be thought of as a special case of the arbitrary data $\{{\bf S}_0,\dots,{\bf S}_N\}$ above) by finding the EMD from $\{{\bf x}_0,\dots,{\bf x}_N\}$ to four different models, $f_j({\bf x}|{\bf y})$, $j=1,2,3,4$.  Each model is of the form in  \ref{gen_model}.  Model $f_1$ has $\alpha=0,\beta=0$, $f_2$ has $\alpha=1.5,\beta=0$, $f_3$ has $\alpha=0,\beta=10$, and for $f_4$, $\alpha=1.5,\beta=10$.  

To show that dharma wheels can detect bias in a process that may not be evident in the underlying model, we simulated 100 animals on a 1000 by 1000 square lattice for 1000 time steps each, performing a random walk with step length distribution $(1/5)\exp[-x/5g(\theta)]$ where $g(\theta)=(8/5)[(9/5)\cos^6(\theta)+(1/5)\sin^6(\theta)]$ and $\theta$ is the animal's bearing.  This means the animal tends to move more in the east-west direction than north-south.  For example, this could be due to a confining valley running from east to west.  We found the EMD between these simulated data and a random walk with a uniform exponential step length distribution, mean 5 units.  We compared it to the EMD between these data and the model from which they were generated.

As a demonstration of the hypothesis test explained in the `Earth mover's distance' section, we use simulated data sets with $\alpha=1.5$ and all integer values of $\beta$ ranging from $0$ to $10$.  We imagine that someone has gathered these data but only knows about, or has data on, Layer 1.  Therefore the best model that this person can construct has $\alpha=1.5$ and $\beta=0$.  We test the hypothesis that this latter model is an accurate description of the various simulated data sets, using the above test.  This mimics a situation where two layers (Layer 1 and Layer 2) are affecting animal movement but the data gatherer has only thought to test one of them (Layer 1).  It tests how well or technique does at informing the user that there is something missing from the model.

For each pair of parameter values $(\alpha,\beta)$, we simulate 500 data sets.  We test the null hypothesis that `a model with $\alpha=1.5$ and $\beta=0$ accurately describes the data' using the above test on each of the 5,500 data sets (500 for each value of $\beta$).  Then, for each value of $\beta$, a certain percentage of the tests accept the null hypothesis, while the rest reject it, so we can plot this percentage against $\beta$ to give a power curve.  A better test would have more hypothesis tests rejected for values of $\beta>0$, whilst having the same or less hypothesis tests rejected for $\beta=0$.  Thus we can assess the relative power of the test using EMD and that using SEMD, by examining the respective areas under the power curves.  We use this to test whether the SEMD improves the power of our hypothesis testing as compared with ordinary EMD.
The power is also likely to be affected by the size of the data set.  To test this, we performed the same power test but for 100 animals moving 1000 steps each.  Since this is highly computationally intensive, we used $M=100$ and simulated only 100 data sets, rather than 500.  Simulations are performed in the C programming language and data analysis in Python.  The Python code has also been translated into R and the C code can be run from R.  The code can be downloaded from 
the Data Dryad Repository (doi:10.5061/dryad.9h42f).  We include an instruction manual in SI Appendix A.

To test the applicability of our technique for models on a real dataset, we use a recent model of bird flock movements and territorial interactions in the Amazon rainforest.  The flocks are multi-species, with the cinerous antshrike ({\it Thamnomanes caesius}) playing a nuclear role in flock cohesion and movement \citep{munn1986}.  Details of the data collection methods, justification for them, the rationale behind the model construction, and the model selection techniques are all given in \citet{PML}.  Though we do not duplicate these specifics here, we give a brief summary of the model.  

The model treats each flock as a single, moving unit.  This reflects the nature of the data which were gathered using the cinerous antshrike's position, where possible, to infer the flock's central location.  The antshrike was usually conspicuous in the centre of the flock.  Data were gathered at 30 second intervals.  The flocks tend to move from tree to tree approximately once every 1-2 minutes.  Model selection techniques reveal that a 1-minute timescale is the best timescale to model these flocks' movement \citep{PMSL} so the model we use has $\tau=1$ minute.  An exponentially decaying distribution of movement lengths between successive 1-minute location fixes is used to model the bird's movement.  The model also includes the intrinsic persistence in the birds' movement.  In addition to this, the birds are modelled as having a preference for higher tree canopies and lower ground.  Finally, the birds know where other flocks have been in the recent past, due to vocalizations, so they are modeled as moving away from areas that have been visited by other flocks.  

We use the EMD testing procedure, given at the end of the previous section, with $M=1000$, to test the hypothesis that the model detailed in \citet{PML} could have given rise to the data observed in the same study.  We examine the resulting dharma wheel as well as the histogrammes of EMD against canopy height, topography, change in canopy height over a step, and change in topography over a step. 

\section*{Results}

\begin{figure}[ht!]
\includegraphics[width=95mm]{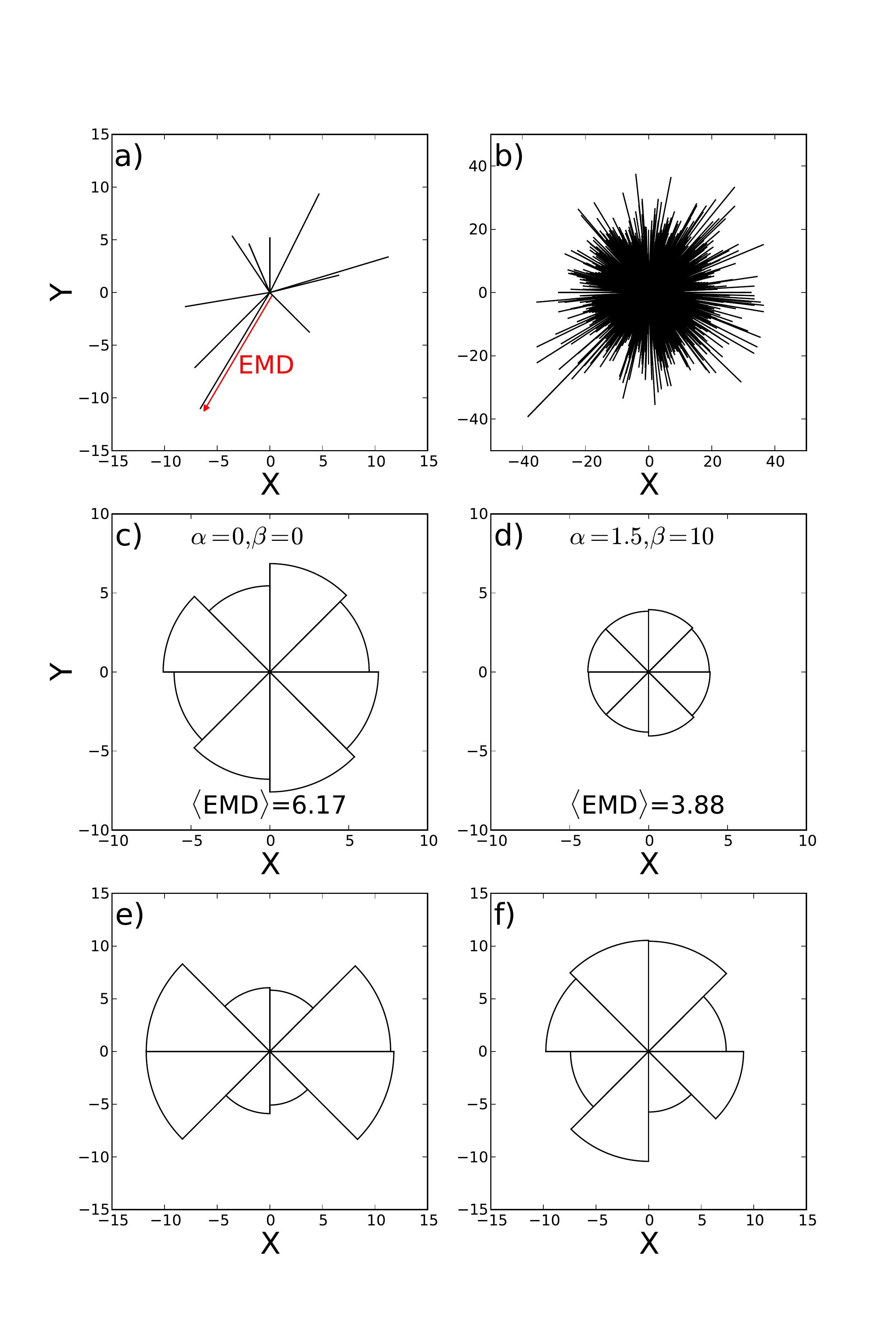}
\caption{\small {\bf Wagon wheels and dharma wheels.}  Panel (a) shows a wagon wheel for a hypothetical data set of 10 points.  The length of each line from the origin is the earth mover's distance (EMD) for a single data point.  The direction of the line is the mean direction from the model to the data point.  This becomes rather messy when there are many data points, as panel (b) shows, where there are 100,000 simulated points.  Instead, we bin the spokes up into eight segments, to construct a dharma wheel (Panels c,d).  The dharma wheels in Panels (c,d) were created using a simulated data set of the model in \ref{gen_model} with $\alpha=1.5$ and $\beta=10$.  The dharma wheel obtained by calculating the EMD from this data set to two different models of the form in \ref{gen_model} are shown here.  The mean EMDs, denoted $\langle$EMD$\rangle$, are given within the panels, together with the parameter values used.  The latter correspond to models $f_1,f_4$ from the main text for panels (c,d) respectively.  Panel (b) was also constructed from $f_1$.  Panels (e,f) were constructed using simulated data from a random walk with a tendency to move faster in the east-west than north-south direction.  Panel (e) shows the dharma wheel using the EMD from this data set to an unbiased random walk model.  Panel (f) shows the EMD to the model from which the data were simulated.}
\label{wheels}
\end{figure}

\begin{figure}[ht!]
\includegraphics[width=150mm]{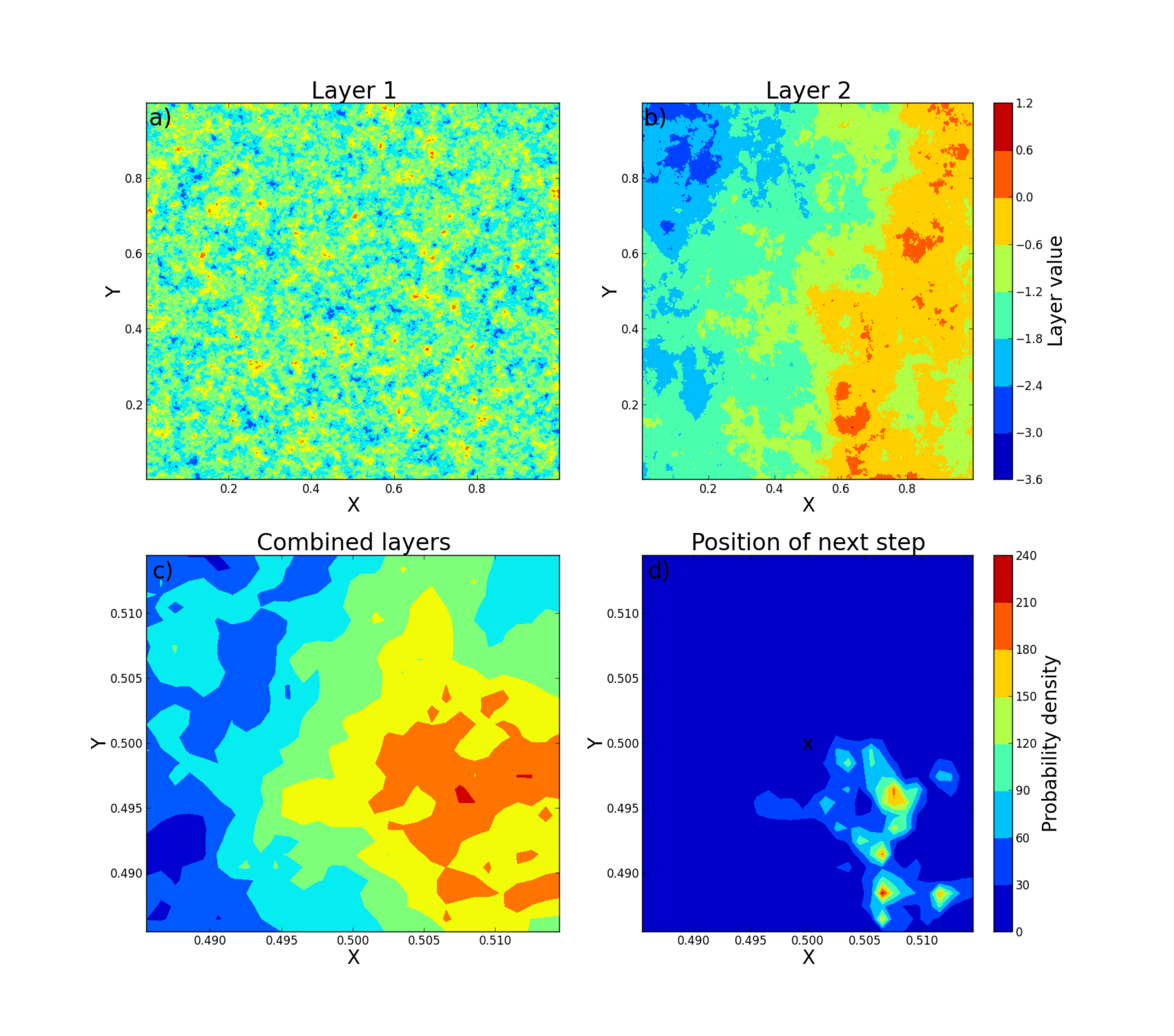}
\caption{\small {\bf Example scenario of a complex movement model.}  An animal moves in a heterogeneous environment, with some randomness but also a tendency to move towards articular regions of space.  Panels (a) and (b) are simulated Geographic Information System (GIS) layers.  The higher the value of the layers at a given point, the more that each movement the animal undergoes is biased towards that point.  Panel (c) shows the combined biasing effect of the two layers in a region close to the center of the simulated study area (using $\alpha=1.5$, $\beta=10$ in the notation of the Methods section).  Panel (d) shows the probability distribution of where an animal, starting at the center, will move to after a time $\tau$ has elapsed (see Methods for details).}
\label{ssf}
\end{figure}

The example process we use is of a hypothetical animal in a heterogeneous environment consisting of two layers that affect movement (Fig.~\ref{ssf}).  As expected, the dharma wheel for the EMD between simulated data and a model that accurately reflects the simulations (Fig.~\ref{wheels}c) has a smaller area than one with certain parameters suppressed (Fig.~\ref{wheels}d, SI figure 2).  Visually, this is not so apparent with SEMD, suggesting that there is value in using EMD for such qualitative tests.

Figs.~\ref{wheels}e,f show what happens when there is a bias in the movement process.  If the model fails to take this into account then the resulting dharma wheel is skewed in the direction of the bias, in this case parallel to the $x$-axis.  A model that does take this bias into account results in a more symmetric dharma wheel (Fig.~\ref{wheels}f), albeit with some random variation.



\begin{figure}[ht!]
\includegraphics[width=150mm]{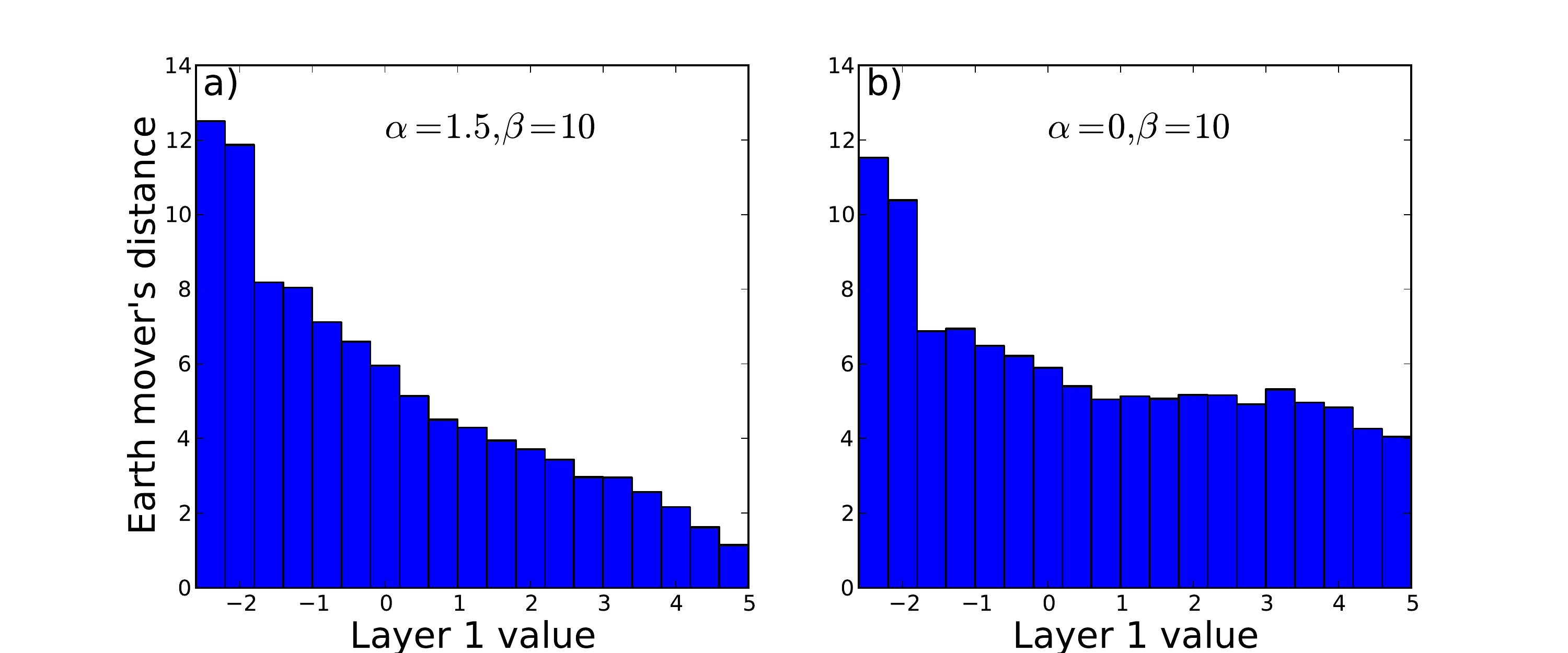}
\caption{\small {\bf EMD binned by value of layer 1.}  Using 100,000 simulated data points from the model \ref{gen_model} with $\alpha=1.5$ and $\beta=10$, the EMDs from these data to the model with (a) $\alpha=1.5$ and $\beta=10$, and (b) $\alpha=0$ and $\beta=10$.  EMDs for each step are binned according to the value of Layer 1 (Fig.~\ref{ssf}a) at the point where the step ends.  Unless this value is very low, the model with $\alpha=1.5$ is better, otherwise a model excluding the effect of Layer 1 is better.  This shows how EMD can be used to ascertain which environments a model may prove to be good at predicting movements, and where it is likely to fail.}
\label{dfr_emd_by_layer}
\end{figure}

By constructing histograms of EMD against the value of the layer where each step ends, Fig.~\ref{dfr_emd_by_layer} shows that model $f_4$ is better at predicting the observations when animals are in environments where the value of layer 1 is relatively high (see Methods for descriptions of models $f_i$).  However, if it is low then model $f_3$ performs marginally better.  This is important if we wish to use a model parametrized in one study area to predict movement in another.  In the example here, if we imagine layer 1 denotes food availability, then Fig.~\ref{dfr_emd_by_layer} tells us that our overall best model, $f_4$, will be good at predicting movement in a food-rich environment but may be quite bad if we try to use it in a food-poor area.

\begin{figure}[ht!]
\includegraphics[width=150mm]{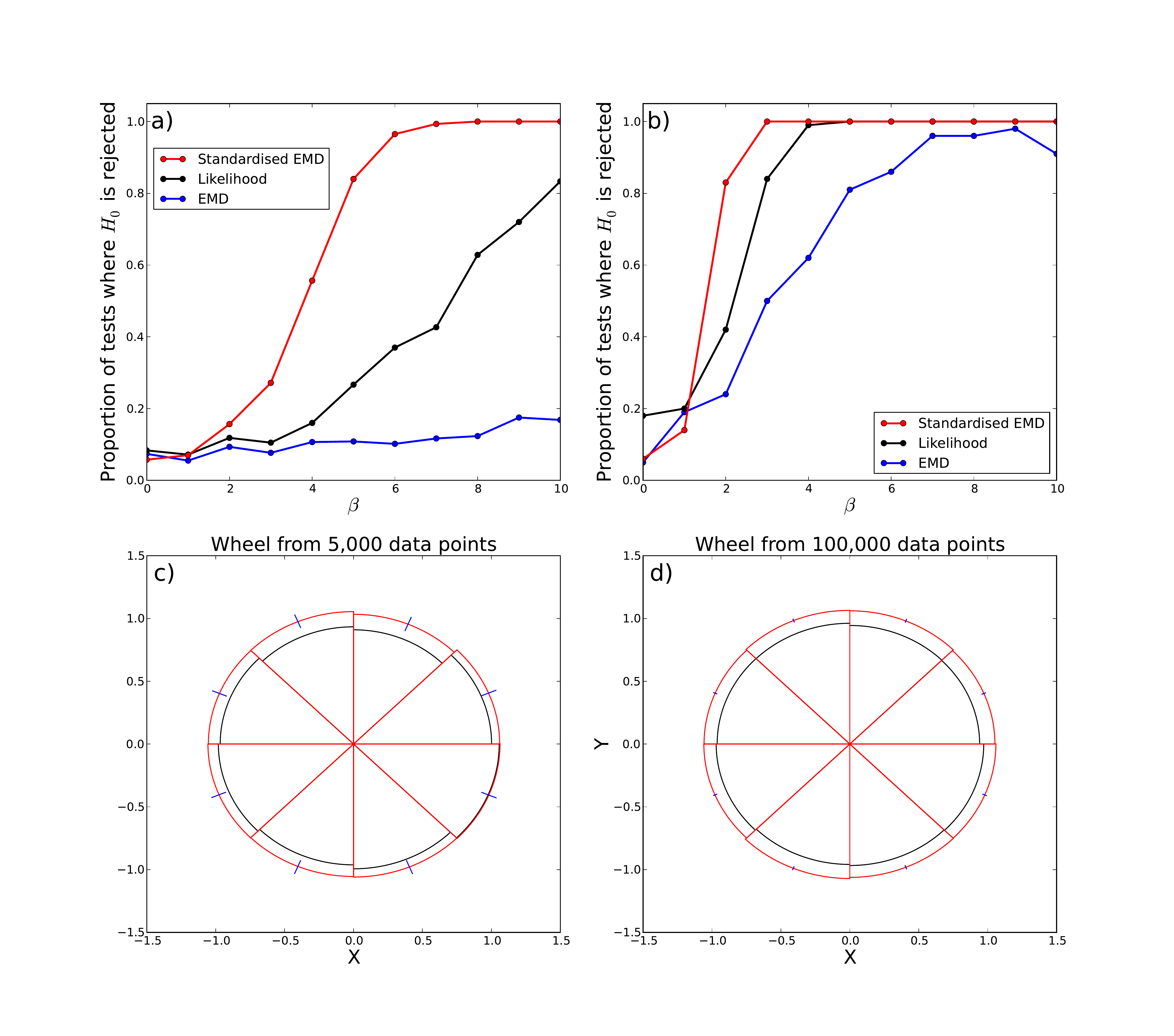}
\caption{\small {\bf Power of the EMD hypothesis test.}  Panel (a) shows the proportion of simulated data sets for which the hypothesis that a model with $\alpha=1.5$ and $\beta=0$ accurately describes the data was rejected, for each value of $\beta$ used.  Using SEMD proves to be far preferable to both ordinary EMD and likelihood.  In all situations, approximately 5\% of data sets with $\beta=0$ result in type I errors, as expected due to the use of 95\% confidence intervals.  As $\beta$ is increased, the number of type II errors decreases, to the point where zero out of 500 data sets exhibited type II errors occurred when using SEMD, if $\beta \geq 8$.  Panel (b) shows a similar plot, but using 100,000 data points for each simulated set, rather than the 5,000 used in panel (a), showing that the larger the data set, the stronger the power of the test.  Panels (c) and (d) represent the hypothesis test visually.  The black curves shows dharma wheels of simulated data with $\alpha=1.5$ and $\beta=10$ tested against a model with $\alpha=1.5$ and $\beta=0$.  The red curves show dharma wheels of the mean of 1000 simulated data sets with $\alpha=1.5$ and $\beta=0$ tested against a model with $\alpha=1.5$ and $\beta=0$ (red curves).  The blue lines show 95\% confidence intervals.  Each simulated data set for constructing Panel (c) had 5,000 points, while Panel (d) used 100,000 points for each data set.}
\label{dfr_power}
\end{figure}

Performing a power analysis of the hypothesis test detailed in the previous section by varying $\beta$ in the simulations (see Methods), SEMD performed far better than ordinary EMD (Fig.~\ref{dfr_power}a,b).  
Ordinary EMD turned out to be quite poor when used for testing the hypothesis that a model accurately reflects the data, almost always failing to reject the null hypothesis when it should, thus being highly susceptible to type II errors \citep{casellaberger2002}.  Replacing SEMD with log-likelihood in the hypothesis test also yielded worse results.  Therefore we would recommend using SEMD for testing this type of hypothesis.

The power of the test depends upon both the actual underlying processes and the size of the data set.  Fig.~\ref{dfr_power} demonstrates that using 100,000 data points (Panels b and d) rather than merely 5,000 (Panel a and c) gives much smaller error bars on the EMD (Panels c and d), and rejects false null hypotheses more frequently (Panels a and b).  

Using the EMD procedure to tested quality of the model of Amazonian bird flock movement from \citet{PML} against the data described there, revealed that the model is insufficient to describe the data in full accuracy.  This rejection occured at both the 5\% and 1\% significance levels.  Using notation from the end of the section `Earth mover's distance: a measure of absolute fit', $\mbox{EMD}_{\rm data}=1.737$, whereas the 0.5 and 99.5 percentiles of $\mbox{EMD}_1,\dots,\mbox{EMD}_M$ are 1.690 and 1.691 respectively.  Using SEMD, the corresponding values are $\mbox{EMD}_{\rm data}=1.153$ and 0.5 and 99.5 percentiles are $1.129$ and	$1.134$ respectively.  

The resulting dharma wheel is very round (SI Figure 3), suggesting that the model is unlikely to be missing a directional bias.  Histogramming the EMDs by canopy height and topography also reveals no clear trend (SI Figure 4a,b).  When we use the difference in canopy height from the start to the end of the step there is also no clear trend in EMD (SI Figure 4c).  However, when we histogram by difference in topography, the EMD is lower for steps where the difference in topography is lower (SI Figure 4d).  This suggests that there may be some additional trigger that causes the particular decisions to move to higher or lower ground.  These observations are unchanged when we use standardized EMD (SI Figure 5).

\section*{Discussion}

We have constructed a generalized version of a residual that is usable with complex stochastic movement models.  We have given a method for using this to test the validity of a model in an absolute rather than relative sense, as well as showing how this concept can give visual insight into the strengths and shortcomings of a stochastic model.  By testing our techniques on simulated data, where we have control over the mechanisms underlying movement decisions, we have demonstrated that our techniques are tractable and can give useful insight into realistic situations.

While many techniques exist for comparing models within a limited class, for example AIC and BIC \citep{burnhamanderson}, our techniques can be used for showing whether there is an important model parameter missing from outside that class.  
This would help mitigate against scientists making bold conclusions after having only fitted data to a small number of poor models, only later to have those conclusions refuted when a more realistic one is used \citep{edwardsetal2007,jansenetal2012,augermetheetal2014}.

Incorporating the correct level of complexity is also of great importance if we want to construct models that are truly predictive, as such models must include all necessary mechanisms to make the predictions accurate \citep{evansetal2013}.  To this end, we believe that combining our EMD techniques with cross-validation may prove to be useful \citep{seymour1993}.  One might use model selection on one subset $\Sigma$ of a data set, then analyze the rest of the data, $\Sigma^c$, by finding the EMD between $\Sigma^c$ and the best model.  If this EMD is very different from the EMD between the best model and $\Sigma$ then this suggests that the model has weak predictive power.  Turning this into a rigorous statistical test would be a useful extension of our approach.

The main aim of our hypothesis test procedure is to evaluate goodness of fit by rejecting models that do not capture the data well.  Though this worked reasonably well in the scenarios we examined, if the data set is too small or the effect of a certain covariate too mild then the procedure can be prone to Type II errors, i.e. failing to reject an incorrect null hypothesis (Fig.~\ref{dfr_power}a). The chance of making Type I errors, i.e. rejecting a correct null hypothesis, is simply the same as the confidence interval used in the testing procedure \citep{casellaberger2002}.  Since we used 95\% in our study, this will happen 5\% of the time, as confirmed by simulations (Fig.~\ref{dfr_power}a).  However, there is no reason in principle why a user of our methods could not use a different confidence interval, or search for the exact $p$-value of the test for their data set, which may be very small if the data set is large.

As well as being a natural generalization of a residual, a strength of our method is that it takes into account multi-modal probability distributions, giving a low EMD to data points that are on any one of many peaks.  Methods such as posterior predictive checks (PPC) \citep{gelmanetal2004} typically examine the mean of a probability distribution, or samples thereof.  Suppose, for example, we have two peaks with a low probability area between them.  Then PPC would penalize an observation on one of the peaks more than in the low area between the two peaks.  EMD, on the other hand, would penalize these observations roughly equally (e.g. Fig~\ref{emm}b).  Therefore one could use EMD as an alternative criterion within a PPC. 
Generalized forms of \ref{EMD}, discussed in SI Appendix B, can have rich properties that allow the user to decide how to penalize areas of a distribution where there are multiple peaks.

Our technique revealed that a recent model of Amazonian bird movement \citep{PML} is insufficient to describe the data it models, inasmuch as we were forced to reject the hypothesis that the data arose purely from processes descibed in the model.  This is despite the fact that the model includes five different factors (step length, turning angle, attraction to high canopies, bias towards lower ground, repulsion away from other flocks' territories) that were all shown in the previous study to have significant impacts on bird movement.  In other words, it is the best model from a variety of different hypothesised models, but it is not good enough for accurate description or prediction of movement.  Consequently, we might be missing a key process in understanding the movement of these animals.  

This corroborates qualitative findings from the previous study, which showed pictorially that this model appeared to give slightly inaccurate territorial patterns when simulated.  However, quantitative confirmation of this is far better than relying on pictures, and it demonstrates that we need to think of further covariates that may be affecting bird movement if we are to build an accurate, predictive model.  Since the dharma wheel for this model is very round (SI Figure 2), it is likely that these covariates do not include directional biases, so we should look for other, non-directional effects.  The reader is referred to the previous study for a discussion of hypothesised candidates for these effects, which include extra confinement due to memory and territorial displays \citep{PML}.

The EMD analysis also revealed that the model does not capture well any moves that flocks make where there is a big difference in topography.  This suggests that there is some additional trigger causing birds to move to higher or lower ground beyond those examined so far.  Flocks display movements akin to patrolling areas on edges of territories on higher ground, possibly to reinforce cues for demarcation of territorial limits.  In such areas, when not engaged in territorial interactions, flocks display more ballistic paths compared to the area restricted behavior seen inside drainage valleys, indicating that intensive area restricted foraging may not be a primary motivator (K. Mokross pers. obs.).  Therefore movements to much higher ground may indicate a switch from foraging to territorial defence, whereas movements to much lower ground may indicate a switch back to foraging.  

Though we have focused on animal movement in this paper, the techniques we propose can, in principle, be used for analyzing any stochastic movement model.  For example, these could include collective motion of humans, cancer growth, or complex systems of intra-cellular proteins \citep{berendsenhayward2000,
friedlwolf2003, helbingetal2007}.  We imagine that the techniques proposed here are merely the tip of the iceberg of possible uses for EMD in analyzing movement models.  We have already suggested some possible extensions, such as using the generalized form in SI Appendix B, or combining EMD with cross-validation.  Analysis of these situations and others, while beyond the scope of the present paper, would doubtless provide further important techniques for understanding how best to model complex systems.  This paper provides an introduction to the concept of residuals generalized to stochastic movement models.  We hope that future work, by us and others, will find many more riches that come from this fundamental idea.

\section*{Acknowledgments}

This study was partly funded by NSERC Discovery and Accelerator grants (MAL, JRP).  MAL also gratefully acknowledges a Canada Research Chair and a Killam Research Fellowship. MAM gratefully acknowledges Alberta Innovates-Technology futures, the Killam trusts, NSERC, and the University of Alberta for graduate student scholarships.  KM would like to acknowledge the Biological Dynamics of Forest Fragments Project (BDFFP) staff for providing logistic support; J. Lopes, E.L. Retroz, P. Hendrigo, A. C. Vilela, A. Nunes, B. Souza, M. Campos for field assistance; M. Cohn-Haft for valuable discussions.  Funding for the research was provided by US National Science Foundation grant LTREB 0545491 awarded to Phil Stouffer, which helped fund KM's work.  This article represents publication no. 648 in the BDFFP Technical Series and contribution no. 34 in the Amazonian Ornithology Technical Series of the INPA Zoological Collections Program. This manuscript was approved for publication by the Director of the Louisiana Agricultural Experiment Station as manuscript 2014-241-16702.  We are grateful to Mike Bryniarski for compiling our code for use with Mac OS, to Phil Stouffer for helpful comments regarding data collection, to members of the Lewis Lab for helpful discussions and suggestions, and for three anonymous reviewers and an Associate Editor for comments that have helped improve the manuscript.

\section*{Data Accessibility}

Data used in this study can be obtained from \citet{pottsetal2014d}.  Code for calculating EMD and generating simulated data can be found from the Data Dryad Repository (doi:10.5061/dryad.9h42f).

\newpage
\section*{Figures}

\end{document}